\documentclass[apj]{emulateapj}

\usepackage{amsmath}
\eqsecnum

\renewcommand\email\texttt

\newcommand\nameobj{Leo V }

\def\spose#1{\hbox to 0pt{#1\hss}}
\def\lta{\mathrel{\spose{\lower 3pt\hbox{$\sim$}}
    \raise 2.0pt\hbox{$<$}}}
\def\gta{\mathrel{\spose{\lower 3pt\hbox{$\sim$}}
    \raise 2.0pt\hbox{$>$}}}

\begin{document} 

\slugcomment{\sc submitted to \it Astrophysical Journal Letters}
\shorttitle{\sc \nameobj Dwarf} 
\shortauthors{}

\title{Leo~V: A Companion of a Companion of the Milky Way Galaxy?}
\author{V.\ Belokurov\altaffilmark{1},
M.\ G.\ Walker\altaffilmark{1}, 
N.\ W.\ Evans\altaffilmark{1}, 
D.\ C.\ Faria\altaffilmark{1},
G.\ Gilmore\altaffilmark{1},
M.\ J.\ Irwin\altaffilmark{1},
S. Koposov\altaffilmark{2},
M.\ Mateo\altaffilmark{3},
E.\ Olszewski\altaffilmark{4},
D.\ B.\ Zucker\altaffilmark{1}
}

\altaffiltext{1}{Institute of Astronomy, University of Cambridge,
Madingley Road, Cambridge CB3 0HA, UK;\email{vasily,walker,nwe@ast.cam.ac.uk}}
\altaffiltext{2}{Max Planck Institute for Astronomy, K\"{o}nigstuhl
17, 69117 Heidelberg, Germany}
\altaffiltext{3}{Department of Astronomy, University of Michigan, 
Ann Arbor, MI 48109, USA}
\altaffiltext{4}{Steward Observatory, University of Arizona, Tucson,
  AZ 85721, USA}

\begin{abstract}
  We report the discovery of a new Milky Way satellite
  in the constellation Leo, identified in data from the Sloan
  Digital Sky Survey. It lies at a distance of $\sim 180$ kpc, and is
  separated by $\lesssim 3^\circ$ from another recent discovery,
  Leo~IV. We present follow-up imaging from the Isaac Newton Telescope
  and spectroscopy from the Hectochelle fiber spectrograph at the
  Multiple Mirror Telescope. Leo~V's heliocentric velocity is $\sim
  173.3 \pm 3.1$ kms$^{-1}$, offset by $\sim 40$ kms$^{-1}$
  from that of Leo~IV. A simple interpretation of the kinematic data
  is that both objects may lie on the same stream, though the implied
  orbit is only modestly eccentric ($e \sim 0.2$)
\end{abstract}


\section{Introduction}

In the last few years, there have been numerous discoveries of
ultra-faint Milky Way satellites, primarily because the Sloan
Digital Sky Survey (SDSS) allows the detection of galaxies with
central surface brightnesses as faint as $30$ mag arcsec$^{-2}$. The
new discoveries include 10 new Milky Way dwarf galaxies, together with
4 unusually extended or faint globular
clusters~\citep{Wi05,Zu06a,Zu06b,Be06a,Be07,Ir07,Wal07,Ko07}.  The
purpose of this {\it Letter} is to announce the discovery of an
additional Milky Way satellite, probably a dwarf galaxy that may be
undergoing disruption, at a heliocentric distance of $\sim
180$ kpc in the constellation of Leo. Following the convention for
naming dwarf spheroidals, we call it Leo~V. It lies very close to one
of our other recent discoveries, namely Leo~IV~\citep{Be07}. Hence, it
is a companion to a companion of the Milky Way Galaxy.

\begin{figure*}[t]
\begin{center}
\includegraphics[width=0.9\textwidth]{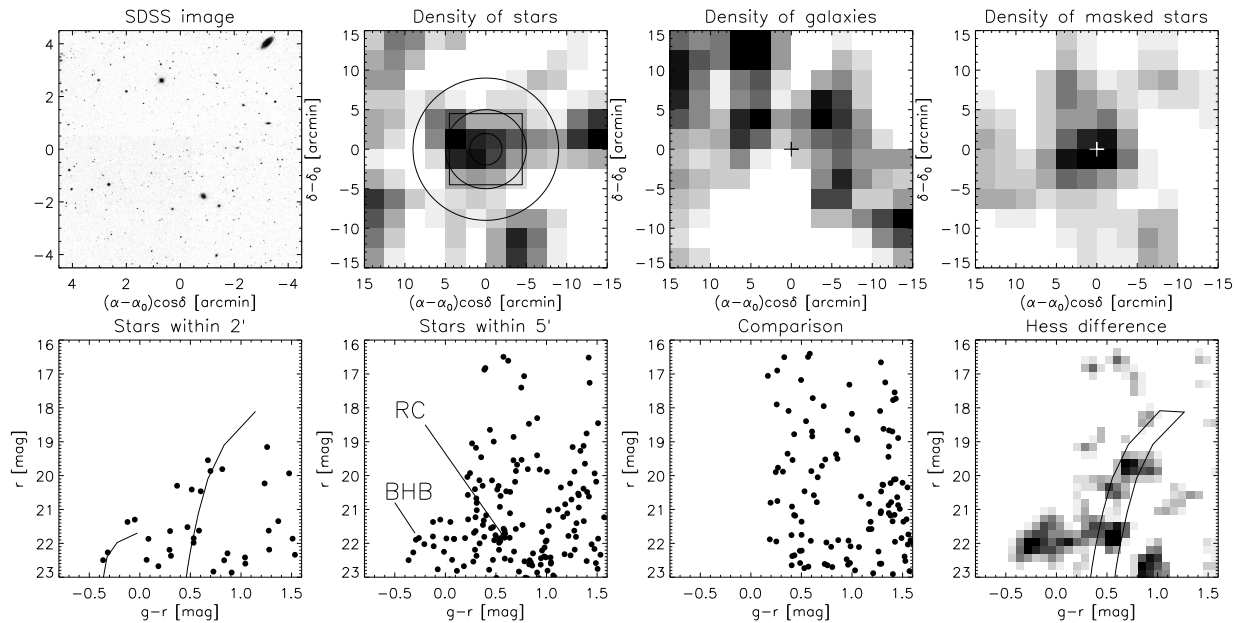}
\caption{The Leo~V Satellite: {\it Upper Left:} SDSS cut-out ($9^\prime
  \times 9^\prime$) around the center of Leo~V. {\it Upper Middle
    Left:} The spatial distribution of all objects classified as stars
  in a $30^\prime \times 30^\prime$ field. {\it Upper Middle Right:}
  The spatial distribution of all objects classified as galaxies. {\it
    Upper Right:} Density of candidate RGB stars selected with the CMD
  mask shown in the panel directly below. {\it Lower Left:} CMD of all
  stars in a circle of radius $2^\prime$, which is expected to be
  dominated by Leo~V members.  {\it Lower Middle Left:} CMD of all
  stars within $5^\prime$, with the red clump and BHB marked. {\it
    Lower Middle Right:} CMD of stars within the annulus $9^\prime$ to
  $10.3^\prime$ showing the foreground. {\it Lower Right:} Difference
  in Hess Diagrams, showing the red giant branch and BHB of Leo~V. The
  mask is built with M92 ridgelines offset to the distance modulus of
  21.25.}
\label{fig:leo_disc}
\end{center}
\end{figure*}
\begin{deluxetable}{lc}
\tablecaption{Properties of the Leo~V Satellite \label{tbl:pars}}
\tablewidth{0pt} \tablehead{ \colhead{Parameter\tablenotemark{a}} &
{~~~ } } \startdata Coordinates (J2000) & 11 31 09.6 +02 13 12.0 \\
Coordinates (Galactic) & $\ell = 261.86^\circ, b = 58.54^\circ$ \\
 $r_h$ (Plummer) & $0\farcm8 \pm 0\farcm1$\\
 $\mu_{\rm 0,V}$ (Plummer) & $27.5 \pm 0\fm5$ \tablenotemark{a}\\
(m$-$M)$_0$ & $21\fm25$\\
 M$_{\rm tot,V}$ & $-4\fm3 \pm 0\fm5$ \\
 $v_\odot$ & $+173.3 \pm 3.1$ kms${}^{-1}$ \\
 $v_{\rm GSR}$ & +60.8 kms${}^{-1}$ 
\enddata
\tablenotetext{a}{Surface brightnesses and integrated magnitudes are
  corrected for the mean Galactic foreground reddening, A$_{\rm V}=
  0\fm1$.}
\label{tab:struct}
\end{deluxetable}

\section{Data and Discovery}

SDSS imaging data are produced in five photometric bands, namely $u$,
$g$, $r$, $i$, and $z$~\citep[see e.g.,][]{Am06,Gu06}. The data are
automatically processed through pipelines to measure photometric and
astrometric properties \citep{Lu99,Sm02,Iv04} and de-reddened using \citet{Sc98}. Data Release 6 (DR6) covers $\sim 8000$ square
degrees, primarily around the North Galactic Pole.

\citet{Ko08} argued that almost all the satellites in SDSS DR5 had
been found and that any further candidates would require substantial
followup imaging to confirm their nature. Accordingly, we
pursued the strategy of acquiring deeper imaging of possible
candidates of lower statistical significance than $6$ (see eq (7) of
\citet{Ko08}). This is of course rather inefficient, and generally
yields negative results.  As the significance is lowered, there are
many candidates that are selected, due to Poisson noise and false
positives induced by large-scale structure. Additional
reasons are thus needed to warrant the expenditure of time and effort. In
the case of Leo~V, its actual significance is $\sim 4$, but the
presence of possible blue horizontal branch (BHB) stars in the SDSS
data was such an indicator.

Fig.~\ref{fig:leo_disc} shows the SDSS view of Leo~V. As
usual with the ultrafaint dwarfs, no object is visible in the SDSS
cut-out. The next two panels of Fig.~\ref{fig:leo_disc} show the
density of resolved stars and galaxies, respectively. There is a
visible overdensity in stars at the location of Leo~V, but the
background shows extensive substructure, leaving the nature of any
object unclear. There is also an overdensity of galaxies close to the
location of Leo~V. Fine-tuning the selection of stars from the
color-magnitude diagram (CMD) to likely members, however, does yield a
convincing object, as shown in the upper right panel of
Fig.~\ref{fig:leo_disc}. The lower panels show three CMDs. The first
is restricted to stars within $2^\prime$ of the center of Leo~V, and
shows a tentative red giant branch (RGB) and a handful of horizontal
branch stars. On moving outwards to stars within $5^\prime$, the
horizontal branch swells and the red clump becomes visible. The
comparison CMD shown in the lower middle right panel is composed of
stars within an annulus of $9^\prime$ to $10.3^\prime$ and shows the
foreground. In the differential Hess diagram in the lower right panel,
there is a convincing detection of the red giant and horizontal
branch. The mask is based on the ridgeline of M92 ([Fe/H] = -2.28,
Clem 2005) and is used to select possible RGB members.

\begin{figure}
\begin{center}
\includegraphics[width=0.5\textwidth]{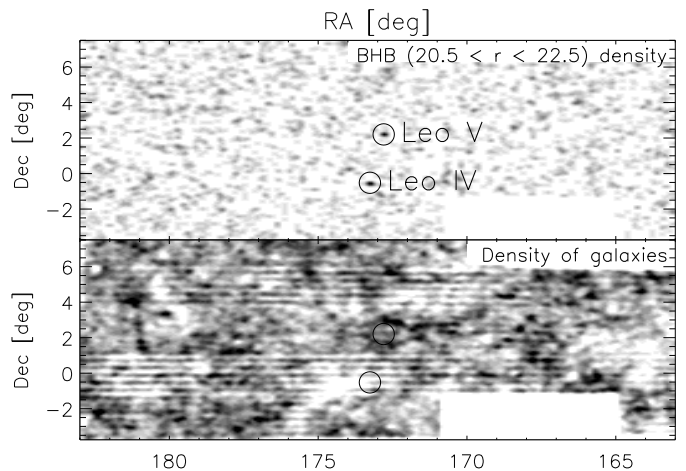}
\caption{{\it Top:} Density of BHB candidate stars in $5^{\prime}$ square
    pixels, smoothed with a $10^\prime$ FWHM filter. Leo~IV and Leo~V are
    marked by circles. {\it Bottom:} Large scale structure at the same
    location.  Note that there is correlation between overdensities in
    the two panels due to object misclassification.}
\label{fig:leobhbdens}
\end{center}
\end{figure}

Fig.~\ref{fig:leobhbdens} is a gray-scale density plot of the BHB stars
selected with the cuts: $20.5 <r <22.5$, $-0.6 < g-r < 0$ and $0.5 < u
-g < 1.5$, based on the cuts of \citet{Si04}. Leo~IV and Leo~V are
clearly visible and separated by only $\sim 2.8^\circ$ on the sky.
There are other overdensities of BHB candidates, but most are
correlated with galaxy clusters as shown in the bottom panel.

\begin{figure*}
\begin{center}
\includegraphics[width=0.9\textwidth]{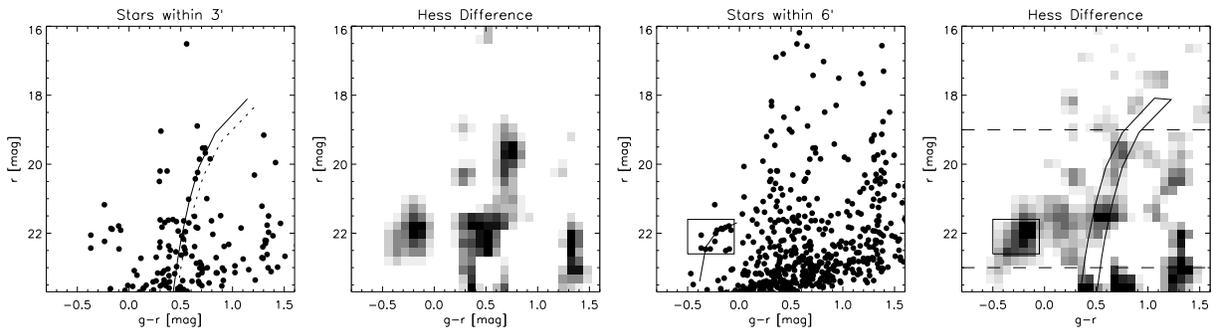}
\caption{Leo~V in INT data: {\it Left and middle left:} CMD and Hess
  differential diagram of stars within $3^\prime$ with ridgelines of
  M92 (solid) and M13 (dotted). Note the possible detection of a red
  horizontal branch. {\it Middle right and right:} The same but for
  stars within $6^\prime$. Note that the BHB stars sit tightly on the
  ridgeline of M92s BHB, offset to the distance modulus of Leo~V.  The
  masks are used to select red giants (within the magnitude range marked
  by the dashed lines) and BHBs.}
\label{fig:leo_int}
\end{center}
\end{figure*}
\begin{figure}
\begin{center}
\includegraphics[width=0.5\textwidth]{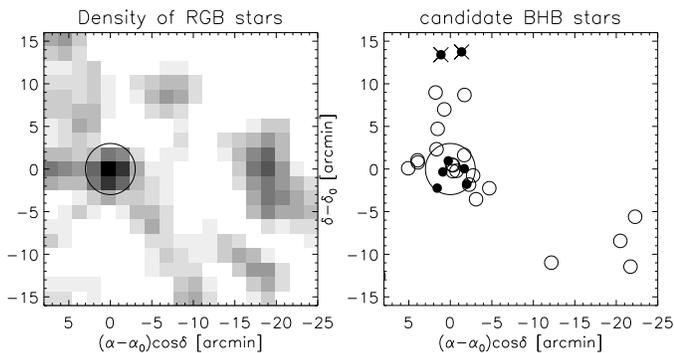}
\caption{{\it Left:} The density of RGB candidate members selected
  from the INT photometry. The extent of Leo~V as judged from two
  half-light radii is marked. {\it Right:} The locations of BHB
  candidate members. Note that the BHB distribution is elongated and
  more extended than that of the RGB stars. Black dots are RGB stars
  with spectroscopy, $v_{\odot} \approx 173$ kms$^{-1}$ and low
  $\Sigma$Mg (see Fig.~5). }
  \label{fig:leo_dens}
\end{center}
\end{figure}
\begin{figure*}
\begin{center}
\includegraphics[width=0.9\textwidth]{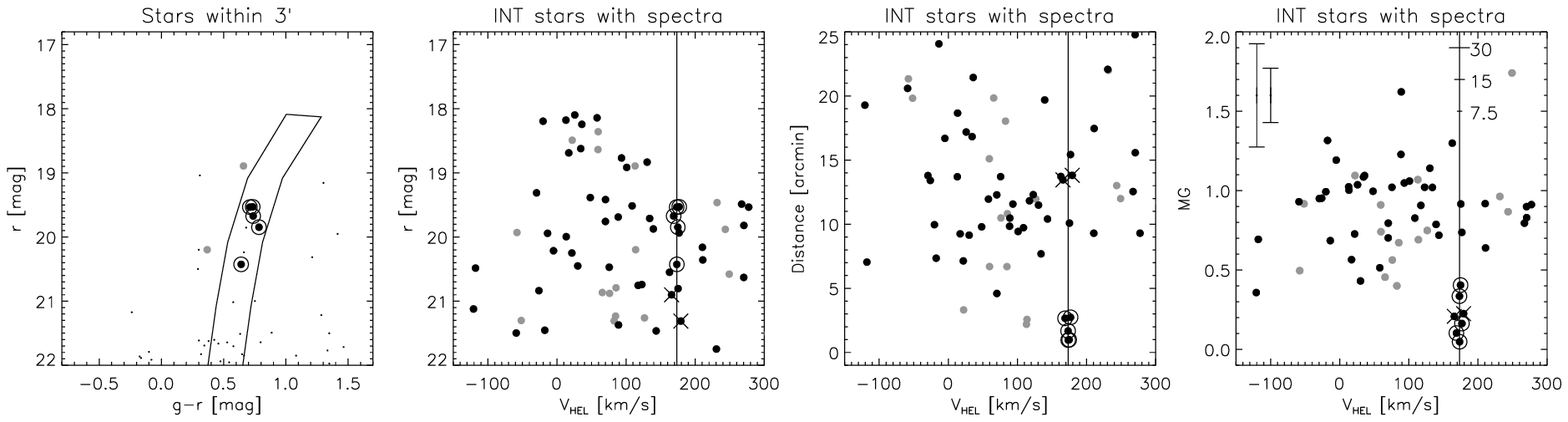}
\caption{{\it Left:} CMD of INT stars within $3^\prime$, shown as
  small dots, together with the mask selecting most likely RGB
  members.  Stars with spectra are full dots, with gray used for stars
  outside the mask and black for those within. The five most probable
  members are circled.  {\it Middle left:} $r$ magnitude versus
  velocity for all INT stars with spectra. The line marks is the
  likely systemic velocity of Leo~V (173.3 kms$^{-1}$). {\it Middle
    right:} Distance from the center as a function of velocity. {\it
    Right:} The pseudo-equivalent width $\Sigma$Mg as a function of
  velocity. The mean and median error in $\Sigma$Mg is shown as
  vertical bars in the left-hand corner. Note the clear separation of
  the giants in Leo~V from Galactic stars. There are two more
  likely members (marked with crosses) located in the data
  cluster. The three horizontal bars have widths corresponding to
  $7.5$, $15$ and $30$ kms${}^{-1}$ to aid calibration.  }
  \label{fig:leo_velocity}
\end{center}
\end{figure*}

\section{Photometric and Spectroscopic Follow-Up}

Follow-up observations of Leo~V were made on 7/8 March 2008 (UT) using
the 2.5 m INT telescope and the WFC mosaic camera, with
four 2k$\times$4k pixel EEV CCDs, a field of view of roughly 30\arcmin
$\times$30\arcmin, and a scale of $0.33^{\prime\prime}$ pixel$^{-1}$ at
the field center. Leo~V was observed with total integrations of 1800s in
$g$ and $r$ filters, split into 3$\times$600s with $\sim
10^{\prime\prime}$ shifts in-between each exposure. The typical seeing
measured directly from the images was rather poor, varying between
$\sim 1.7 - 2.0^{\prime\prime}$. Data were reduced using a general
purpose pipeline for processing wide-field optical CCD
data~\citep{Ir01}. Images were de-biased, trimmed, cross-talk
corrected, and then flatfielded and gain-corrected to a common
internal system using clipped median stacks of nightly twilight flats.
For each image frame, an object catalog was generated and used to
update the world coordinate system prior to stacking each set of 3
frames.  A final set of object catalogs were generated from the
stacked images and objects were morphologically classified as stellar
or non-stellar (or noise-like).  The detected objects in each passband
were then merged by positional coincidence (within $1^{\prime\prime}$)
to form a combined $g,r$ catalog and photometrically calibrated on the
SDSS system using stars in common.

With the poorer than average seeing, the INT data are only about $0.5$
magnitude deeper than the SDSS data. Fig.~\ref{fig:leo_int} shows the
CMDs of stars within $3^{\prime}$ and $6^{\prime}$ from the
center. There is some improvement in the tightness of the red giant
branch and especially the horizontal branch. The BHBs are now aligned
with the ridgeline derived from M92, which is used to measure the
distance to Leo~V as $180 \pm 10$ kpc. Having fixed the distance, we
can experiment with different stellar populations, shown in the first
panel by the ridgelines of M92 (solid) and M13 (dotted). M13 ([Fe/H] = -1.54)
has a giant branch that is too red,
while the more metal-poor M92 ([Fe/H] = -2.28) is a closer match to
the stellar population. We also show masks wrapped around the red
giant and horizontal branches which are used to select the candidate
members for Fig.~\ref{fig:leo_dens}. The two populations are
distributed differently; most of the light from the RGB stars
is limited to the inner $3^{\prime}$, whereas the BHB stars extend out
to at least $10^{\prime}$. This phenomenon has been seen in other
dSphs such as Carina and Sculptor~\citep{Ha01,Tol04,Ko06}


The number density of stars defined by the RGB and BHB selection boxes
is sharply peaked in the central region with a half-light radius, from
Plummer and exponential model fits, of $\sim$0.8$^{\prime}$, or 42
pc for a distance of 180 kpc.  However, the profile also shows an
extended plume of stars slightly above the general background level (Fig.~\ref{fig:leo_dens}).  
This extended appearance makes the
luminosity of the satellite difficult to estimate directly. We first
converted the number density radial profile to a (luminosity-weighted)
surface brightness profile to directly estimate the central surface
brightness.  Integrating the Plummer law model fit then gives a total
flux from resolved stars.  Comparison with the M92 luminosity function
suggests we are missing roughly 1/2 of the light from fainter members
which would yield a total magnitude in the central $3^{\prime}$ radius
region of M$_{\rm tot,V} \approx -4\fm3$. This number is a lower limit
on the luminosity, as it ignores any contribution at larger radius,
such as from the plume.

We obtained spectra of 159 red giant candidates using the Hectochelle
fiber spectrograph at the MMT 6.5-m telescope on Mt. Hopkins, Arizona.
Spectroscopic targets were selected from the red giant branch of Leo V
(Figure~\ref{fig:leo_velocity}) within a field of radius 30$\arcmin$,
centered on the object.  The Hectochelle spectra sample at high
resolution ($R \sim 25000$) the wavelength range $5150-5300$ \AA,
which includes the prominent magnesium triplet (MgT) absorption
feature.  For each of two distinct Hectochelle configurations
targeting Leo V, we obtained $3\times 2700$s exposures during the
nights of 28 and 29 May 2008.  Spectra were reduced following a
procedure described by \citet{Ma08}.  For each star we measure the
(solar rest frame) line-of-sight velocity, $v_\odot$, by
cross-correlating the spectrum against a high-S/N template of known
velocity.  We also measure the pseudo-equivalent width of
the MgT feature, $\Sigma$Mg,  using the technique of \citet{Wa07}.
The data include 70 stars with INT photometry, of which 52 lie inside
the mask shown in the left panel of Fig.~\ref{fig:leo_velocity}.  We
calculate errors in $v_\odot$ and $\Sigma$Mg from models which consider
the quality of the cross-correlation function and spectral S/N,
respectively (see \citealt{Ma08, Wa07}).

The left panel of Figure~\ref{fig:leo_velocity} shows the CMD of the
INT stars with the candidate selection mask superimposed. This is
slightly broader than the one shown in the rightmost panel of
Fig.~\ref{fig:leo_int}, to include all possible candidates. Solid dots
show stars with spectra, gray lie outside the mask and black lie
inside. We circle the five most probable Leo~V members, which have a
mean velocity of 173.3 $\pm$ 3.1 kms$^{-1}$. The remaining three
panels show the correlation between $r$ magnitude, distance from
center and $\Sigma$Mg. Note in particular that in the plane of
($v_\odot, \Sigma$Mg), the giants in Leo~V are clearly separated from
the dwarfs in the thick disk and halo of the Milky Way.  It is also
clear that there are 2 more possible members that lie at large
distance from the center.


\section{Discussion and Conclusions}

Leo~V is most probably a new dwarf galaxy, based on the presence of
an old, metal-poor population with a characteristic size of between 50
and 200 pc. There are several hints that it may be a disrupting
satellite, but our data do not support the idea that it is merely an
overdensity in a stellar stream.  The most remarkable feature of
Leo~V is the disparity in the spatial extent of the RGB and BHB
populations. The red giants are confined to a tight core of $\sim 50$
pc, whereas the BHBs extend out at least as far as 200 pc. There even
appear to be BHBs associated with Leo~V at distances of 500 pc. One
possible explanation for the apparent difference is that the RGB stars
do follow the BHBs much further out, perhaps because the object is
disintegrating, but they are more difficult to distinguish from the
foreground populations in our data.  A hint that this the case is
perhaps provided by the two RGB candidates that are even further than
the most distant BHBs, although their membership needs to be
confirmed.  With deeper photometry reaching down to the turn-off, it
should be possible to verify this hypothesis.  Another explanation is
that the RGB and BHB populations probe different epochs of star
formation in an ultra-low mass system.

Leo~V's possible association with Leo~IV is also unique. Although
there are other examples of dSphs separated on the sky by a few
degrees -- such as CVn~I and CVn~II -- they are at different distances
and velocities. By contrast, Leo~IV is at a heliocentric distance of
$\sim 160$ kpc~\citep{Be07} and a heliocentric velocity of $132$
kms$^{-1}$~\citep{Si07}. These are very close to our estimates of
$\sim 180$ kpc and $173$ kms$^{-1}$ for the distance and velocity of
Leo~V. Referred to the Galactic Standard of Rest, the velocities of
Leo~IV and V are low, $11.0$ kms$^{-1}$ and $59.5$ kms$^{-1}$
respectively.  Using the velocity distribution for the $\rho \sim
r^{-3.5}$ radial profiles given in \citet{Ev97} to construct
artificial samples of 50 satellites, we estimate that there is a
$\lesssim 1$ per cent probability of this coincidence happening by
chance. We remark that one of BHB stars considered by \citet{Si07} for
possible membership of Leo~IV, but then discarded, has a heliocentric
velocity of $160$ kms$^{-1}$. This hints at the possible existence of
extended stellar structures around Leo~IV and Leo~V. If Leo~IV and
Leo~V are assumed to be on the same stream, then the orbit can be
computed, assuming a singular isothermal sphere with amplitude $v_0$
= 220 kms$^{-1}$. The pericenter is $\sim 160$ kpc and the apocenter
is $\sim 244$, so that the eccentricity is modest ($e = 0.2$) and the
orbit never approaches the inner parts of the Milky Way (in which case
both objects should be relatively intact, at odds with their seemingly
irregular appearances).

Leo~V may prove to be an important object for testing theories of
galaxy formation. \citet{Ri08} has argued that very old dwarf galaxies
must form preferentially in chain structures, tracing the filamentary
dark matter in the early universe. These chains or groups of dwarfs
may retain some of their integrity even on accretion and merging into
the Milky Way halo. Is it possible that Leo~IV and Leo~V are two links
in such a chain?

\vskip 0.2truecm
\acknowledgments 

Funding for the SDSS and SDSS-II has been provided by the Alfred P.
Sloan Foundation, the Participating Institutions, the National Science
Foundation, the U.S. Department of Energy, the National Aeronautics
and Space Administration, the Japanese Monbukagakusho, the Max Planck
Society, and the Higher Education Funding Council for England. The
SDSS Web Site is http://www.sdss.org/.  EO acknowledges NSF grants
AST-0205790, 0505711, and 0807498; MM acknowledges NSF grants
AST-0206081 0507453, and 0808043


\end{document}